# Antenna Technology Readiness for the Black Hole Explorer (BHEX) Mission

T. K. Sridharan(1), Robert Lehmensiek(1), Sebastian Schwarz(2), and Daniel M. Marrone(3)
(1) National Radio Astronomy Observatory, Charlottsville, VA, USA
(2) Airbus Defence and Space GmbH, Immenstaad, Germany,
(3) University of Arizona, Steward Observatory, Tucson, AZ, USA

## Abstract

The Black Hole Explorer (BHEX) will be the first sub-mm wavelength Space Very-Long-Baseline Interferometry (VLBI) mission. It targets astronomical imaging with the highest ever spatial resolution to enable detection of the photon ring image feature of a supermassive black hole. BHEX is being proposed for launch in 2031 as a NASA Small Explorers mission.

The BHEX science goals and the mission opportunity require a large high precision lightweight spaceborne antenna. A survey of the technology landscape for realizing such an antenna is presented. Technology readiness (TRL) for the antenna is discussed and assessed to be at TRL 5. A description of our technology maturation efforts towards TRL 6 is provided. Design studies leading to the preliminary design of a high efficiency antenna based on metallized carbon fiber reinforced polymer (CFRP) technology with a 3.4 m primary reflector, a surface precision of < 40 µm to allow efficient operation up to 320 GHz, and a mass of only ~ 50 kg are outlined. Current plans anticipate attaining TRL 6 in 2027-28. Completed studies leading to the preliminary design point to a large margin in surface precision which also opens up applications at significantly higher THz frequencies, beyond BHEX.

## 1. Introduction

The Black Hole Explorer (BHEX) will be the first sub-mm wavelength space-VLBI mission. It aims to discover and measure the finely structured *photon ring* feature that is predicted to exist in images of black holes, produced from light escaping to the observer after orbiting the black hole [1]. BHEX will deliver images with a ~ 6 $\mu$arcsec spatial resolution, the finest ever in astronomy. This is achieved by coherent combination of electromagnetic signals collected separately at sensitive ground antennas and the orbiting BHEX space antenna to realize the resolving and imaging power of long space-ground baselines, corresponding to a telescope of the size of the orbit (~30,000 km), through VLBI techniques. Through these capabilities, BHEX enables pathbreaking advances in fundamental physics and astrophysics. The BHEX mission concept, which is being proposed as a NASA Small Explorers (SMEX) mission by a multi-institution collaboration led by the Center for Astrophysics | Harvard and Smithsonian [2] is outlined in Fig 1. The BHEX science objectives require high-sensitivity observations at mm/sub-mm wavelengths, placing stringent constraints on the antenna.

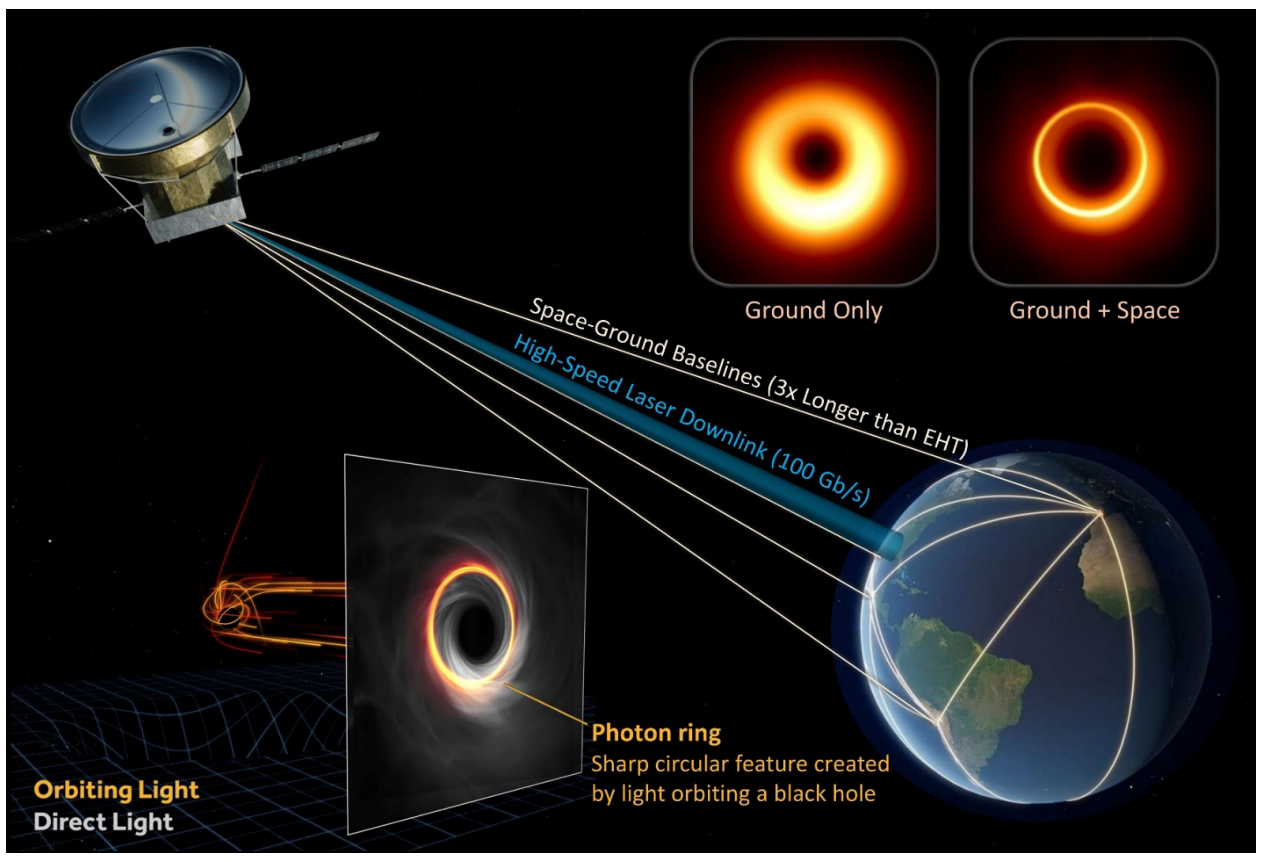

**Figure 1**. BHEX mission overview [1]

We present an assessment of the technologies available for the construction of the antenna and discuss its mechanical design. The development of the optical and antenna designs leading to the current baseline and an overview of the mission instruments have been previously presented [3, 4, 5, 6].

## 2. Antenna Requirements

The high sensitivity requirement translates to the largest possible aperture to maximize collecting area, a high efficiency optical design, high surface precision maintained under a range of in-orbit thermal conditions to increase the surface precision component of the aperture efficiency (low Ruze loss), and low SWaP (size, weight, and power) metrics that can fit within the SMEX mission profile. The displaced axis dual shaped antenna optical configuration and the design parameters chosen maximize the non-Ruze components of the aperture efficiency [3]. Additionally, manufacturability is an important consideration.

For a VLBI instrument, the final interferometric radiation pattern is computationally synthesized on the ground by cross correlating signals coherently recorded at individual stations. Therefore, station primary beam pattern is not important with peak gain being the key metric. Accordingly, as an on-axis design increases the projected effective area for a given physical area, offset designs such as the one employed in the Planck mission to realize a clean beam were discarded.

Axially symmetric antennas are also generally easier to accommodate with launchers from a mechanical standpoint and with downstream optics and receiver systems. Non-circular apertures were also considered to leverage the aspect ratio of payload fairings to accommodate a larger aperture but eliminated as their aperture efficiencies were found to be considerably lower than for axially symmetric systems.

**Table 1** Antenna requirements

| | |
|---|---|
| Configuration | Axially symmetric |
| Diameter | 3.4 m |
| Surface RMS<br>$\eta_{Ruze}$ (sensitivity loss) | < 40 $\mu$m<br>86 GHz : > 98% ( 1%)<br>230 GHz : > 85% ( 7%)<br>320 GHz : > 75% (13%) |
| Blockage and illumination | < 20% loss |
| Mass | < 50 kg |
| Thermal management | 90° sun avoidance, heating and sun shielding |

To maintain the high surface precision in operational conditions, direct sun incidence on the primary reflector (PR) surface was excluded by instituting a sun avoidance zone and shielding the backsides of the PR and the sub-reflector (SR). Table 1 presents the subsystem level baseline antenna requirements identified by the instrument team to accommodate these constraints. Since the space borne antenna constitutes one element of an interferometer formed with a ground antenna, the sensitivity loss is ~ half the power loss.

## 3. Current Antenna Technologies

The combination of the large aperture (3.4 m), the required operating frequency range of 80 - 320 GHz (up to < 1 mm wavelength), and therefore the surface precision, and low mass, falls in a gap in the large spaceborne antenna technology landscape. Lightweight antennas with unfurled/deployable metal mesh fabric surfaces are typical in the X band, e.g. communication and synthetic aperture radar (SAR) applications. Although the mesh technology is feasible in Ka band and offered up to V band, carbon fiber reinforced polymer (CFRP) sandwich reflectors are the choice for high gain antennas (HGA) at higher frequencies, e.g., the Europa Clipper and the Roman Space Telescope communication HGAs and main reflectors of the Planck telescope and the EarthCARE CPR (Cloud Profiling Radar). As CFRP is not a good reflector at higher frequencies, the reflecting surface requires metallization. Heavy mirrors become necessary in the THz and Far Infrared (FIR) bands to achieve large apertures with the required high surface precision, e.g., Herschel.

We carried out a survey of available technologies to identify the most appropriate path for the BHEX antenna which operates in the intervening mm/sub-mm wavelength range, with a large aperture. Representative examples of available technologies and realizations along with their TRLs (see section 4) are shown in Table 2 (not exhaustive), which combines available information from the literature, vendors and our estimates. Only examples with relevant parameters in the range of interest for BHEX are shown. The areal density (mass/surface area) is a key determinant of the final antenna mass and is listed where available.

A public Request for Information (RFI) process led by NRAO elicited multiple submissions from leading vendors in the space astronomy missions area which were evaluated resulting in the selection of Airbus Defence and Space, GmbH (ADS) as the BHEX antenna provider. ADS has strong heritage in the area through the Planck and Herschel missions (Herschel not listed due to its heavy silicon carbide mirror).

## 4. Technology Readiness

NASA has defined a technology readiness level (TRL) classification framework and an associated TRL scale that measures the maturity of a technology. TRL 1 corresponds to the identification of the basic concepts and principles underlying a technology, progressively developed to culminate in a successful operational space mission designated as TRL 9 [7]. This methodology is adopted for BHEX antenna technology assessment.

Antenna manufacture using CFRP sandwich construction is a well-established technology in the space communication high gain antenna (HGA) arena with numerous units flown. Thus, the basic CFRP sandwich antenna technology is TRL 9. The BHEX antenna provider ADS has long standing experience in this area. To estimate TRL for the BHEX antenna, its specific parameters in relation to the heritage examples need to be considered carefully.

The first CFRP antenna/reflector examples used in astronomy missions are the 1.1 m ODIN mission antenna (2001) and the ADS manufactured 1.6 × 1.91 m Planck reflector (2009), both delivering very high surface accuracies of ~ 10 $\mu$m rms. The four CFRP TRL 9 examples (Table 2) with parameters closest to the BHEX antenna target range are the Planck primary reflector (2009), the Europa Clipper HGA (2024), the EarthCare CPR antenna which ADS integrated (2024) and the Roman HGA (integrated 2023; 2027 flight).

A key element of TRL assessment is establishing that the requirements are bounded by the heritage examples for all parameters. This is clearly the case for the ensemble of the examples. The Planck antenna bounds the BHEX antenna surface accuracy by a large margin (×3), is ~ 50% in scale, and employed VDA metallization, but exceeds the BHEX areal mass density target by > ×2. The 3-m Europa Clipper antenna bounds the scale to within 15% at an areal mass density better than the BHEX target but exceeds the surface profile accuracy bound. The 2.5-m EarthCare antenna is within 33% of the BHEX surface accuracy target at 70% scale with an undisclosed areal mass density. The Roman HGA has the same areal mass density as the BHEX target, is at 50% scale, with a poorer surface accuracy by a factor of ~ 2. On any individual TRL 9 example, the parameters are close to the BHEX requirements. Following NASA TRL best practices guide [7], this heritage establishes TRL 5 for the BHEX antenna, where the size scale is not critically important.

**Table 2** A summary of available antenna technologies

| Mission | TRL | Vendor | Size (meter) | Surface accuracy (micron) | Areal density (kg/m$^{2)}$ | Operating Band | Technology |
|---|---|---|---|---|---|---|---|
| Planck | 9 | Astrium/ADS | 1.6 ×1.9 | 10-50 | 13 | mm-THz | Metallized CFRP sandwich |
| BHEX target | 5* | ADS | 3.5 | < 40 | ~ 5 | mm/sub-mm | Metallized CFRP sandwich |
| EarthCare CPR | 9 | NEC | 2.5 | < 60 | - † | mm | CFRP sandwich |
| Europa Clipper HGA | 9 | AASC | 3 | < 150 | 4 | Ka | CFRP sandwich |
| Roman HGA | 9 | AASC | 1.8 | < 85 | 5 | Ka | CFRP sandwich |

* as assessed in this work; † this information is not available; ADS: Airbus Defence and Space GmbH, Germany; NEC: NEC Corp., Japan; AASC: Applied Aerospace Structures Corp., USA

## 5. Technology Maturation

We take a cautious approach to technology maturation for gaining the small expansion of the parameter bound envelope required for the BHEX antenna over the individual heritage examples and to advance its TRL beyond TRL 5. In planning this effort, we closely adhere to the NASA best practices guide for TRL assessment [7]. TRL 6 is formally defined as building a high-fidelity prototype of the system/subsystems that adequately addresses all critical scaling issues and testing it in a relevant environment to demonstrate performance under critical environmental conditions.

As BHEX has adopted a proto flight model (PFM) philosophy to limit costs, realizing a separate full scale model early on is not an option. Therefore, following the NASA guide, the path chosen to establish TRL 6 is through high fidelity FEM/thermal modeling analysis of the full-scale design using known material properties and worst case operational conditions, and the modeling, fabrication, and targeted testing of a half-scale demonstrator model (DM), allowing correlation of the analyses and test results to gain high confidence in the predicted full scale performance.

## 6. Preliminary Design

As the first step, ADS has completed the development of the preliminary design of the full scale antenna and has carried out modeling and analysis to quantify its expected performance in a range of operational conditions encompassing the worst case scenarios (renditions shown in Fig 2).

To achieve a lightweight, stiff and stable structure, the PR is designed completely from a CFRP as a single sandwich and stiffened with CFRP sandwich backing structures. The small SR is made of aluminum and is supported by CFRP struts. The optical design positions the SR and the struts inside the PR rim [3] which excludes sun incidence, leading to adequate strut thermal management with single layer insulation (SLI). The backside of the SR is shielded by a multi-layer insulation (MLI) cap. The interface of the antenna to the spacecraft is through a 3×bipod CFRP kinematic mount. The entire PR and the mount are enclosed in an MLI skirt blanket for thermal management. The thermal management concept further includes channeling of spacecraft heat dissipation to radiatively heat the PR and actively controlled electrical heating of the SR. The baseline PR metallization concept is electroless chemical silver deposition with a gold overcoat.

An essential set of trades were conducted as part of the design development. Multiple backing structure architectures - iso grid vs spider and different spider configurations - were studied through stiffness and mass analyses leading to the selection of the configuration shown in Fig 2. Analyses of multiple configurations for the antenna to spacecraft interface identified the 3×bipods structure as optimal through modal and deflection analyses. The chosen iso-static configuration shown in Fig 2 isolates spacecraft interface induced deflections better. The SR support struts directly interface to the PR mount points for load transfer. Single tube SR strut structure was adopted due to lower blockage than bipods [3].

The thermal and mechanical properties of the materials that go into the design are well known from heritage projects, allowing robust modeling to predict the performance of the antenna. This modeling includes detailed thermal analysis in BHEX orbital and attitude scenarios derived from science requirements and two spacecraft heating cases (50 W and 200W) to determine realistic temperature distributions, which were then used as inputs to the following thermo-elastic distortion (TED) analysis. Two of the extreme operational thermal load cases analyzed are shown in Fig 3, where (a) BHEX is in eclipse which results in the coldest conditions and (b) BHEX points close to the earth's limb with the PR receiving full earth shine and full sun incidence from the back, leading to the hottest conditions. The results of the TED analysis for the worst-case thermal load conditions are combined with other influence factors, viz., gravity release, moisture release and satellite imposed deflections and alignment errors to arrive at the worst case antenna deflection. This analysis has shown that the antenna does not run into any critical situation and temperature limits of the materials and design are not exceeded. The resulting worst case reflector surface error is 16 $\mu$m, well within the 40 $\mu$m specification, leaving a large margin to realize higher efficiencies. The total mass of the antenna meets the < 50 kg specification. The results point to opportunities for applications beyond BHEX, at significantly higher THz frequencies.

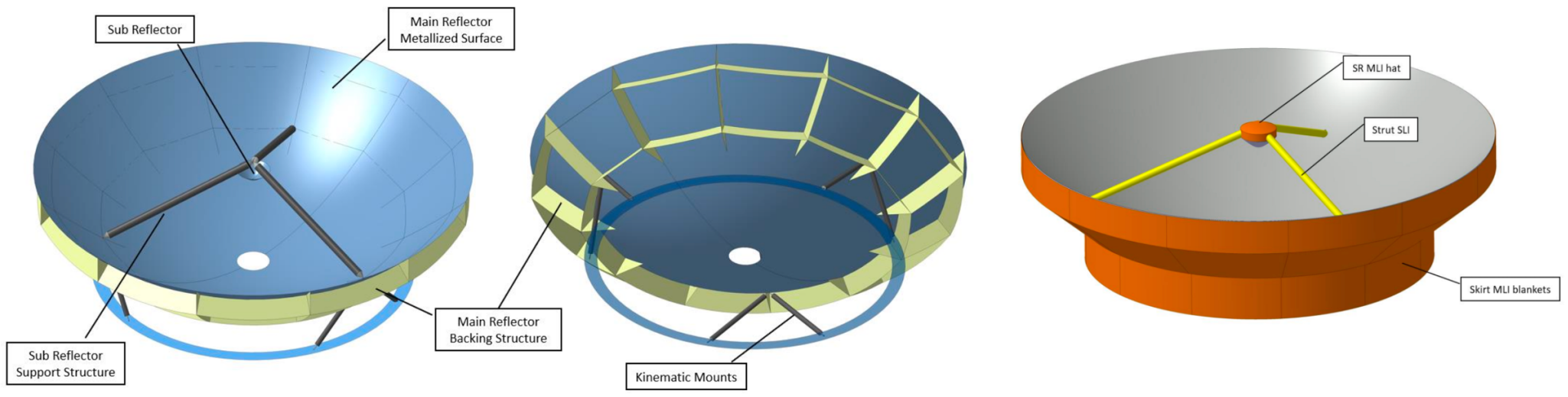

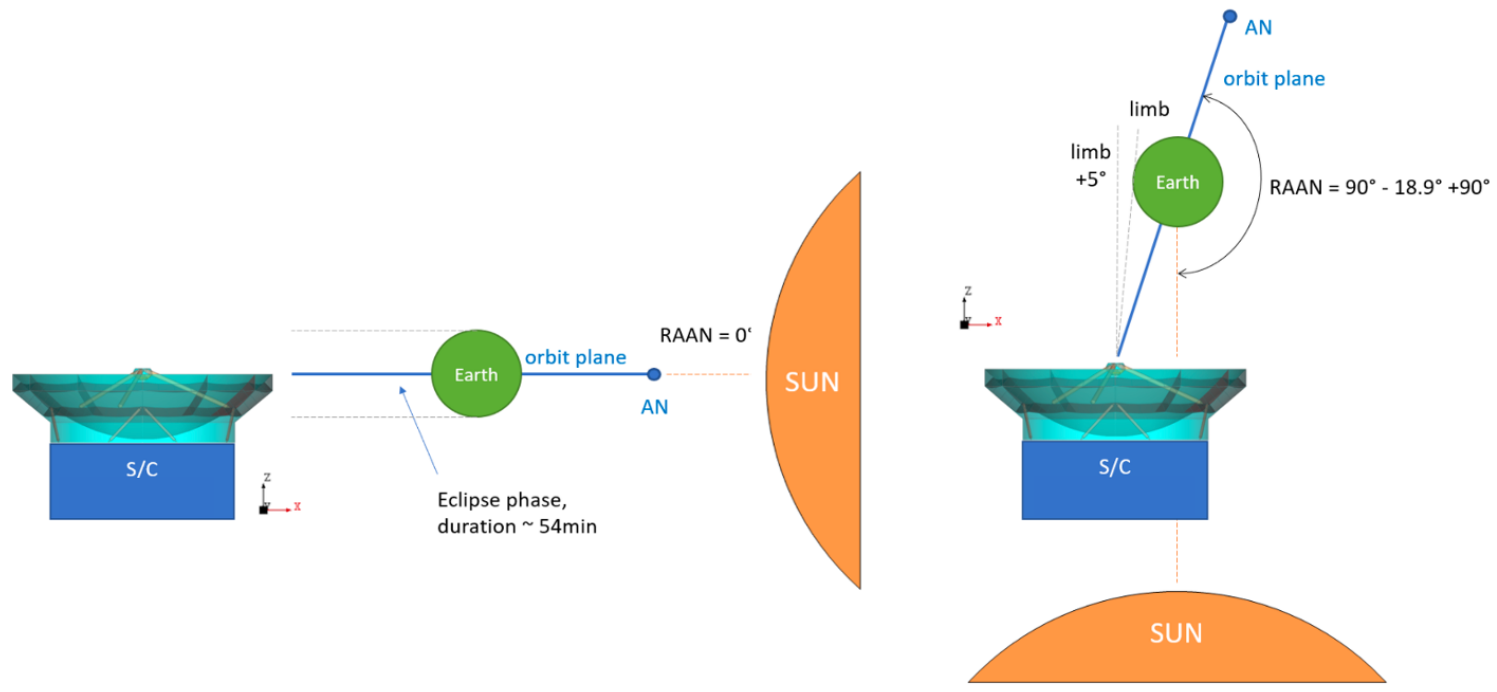

**Figure 2 & 3:** A rendition of the preliminary design and extreme operational cases modeled.

## 7. Current Activities and Next Steps

The BHEX antenna is being developed at ADS in a common program with pre-development for the ESA selected WIVERN mission antenna which has comparable parameters. Under this program, ADS has fabricated a half-scale PR sandwich meeting the areal density target incorporating the same design principles and materials as the full scale BHEX design. Metallization samples show excellent RF reflectivity test results in the 3-mm band, with 320 GHz tests planned.

The BHEX technology maturation program (section 5) under way targets TRL 6 in 2027-28, adding CFRP back up structure to complete the PR, followed by model results validation (section 5 & 6) through tests and measurements. This includes measuring manufacturing errors (i.e. the mold to CFRP DM replication errors) and space qualification tests.

FEM analysis predictions from the full-scale design and modeling (section 7) would guide the tests of the half-scale DM, by specifying test parameters and critical load and stress monitoring points during sine and random vibration tests. The thermoelastic deformations (TED) of the DM will be measured over a temperature range exceeding the operational extremes using a metrology assembly in a thermal vacuum (TVAC) chamber and compared with the model results: profile measurements after initial bake out and during thermal cycling. Successful completion of these qualification tests of the DM would establish scaling and performance in relevant environment under critical conditions, satisfying the requirements for TRL 6.

## 7. Conclusions

We have conducted a survey of the technologies available for manufacturing the large aperture, high precision, lightweight antenna for the BHEX sub-mm wave space-VLBI mission and identified metallized CFRP technology as the appropriate choice. The current maturity level of the technology is assessed to be at TRL 5. A preliminary design for the antenna has been developed through mission level choices, a preliminary high efficiency optical design, structural and thermal modeling and analyses, and a set of trades. Performance modeling results under worst case thermal conditions comfortably meet the BHEX antenna specifications. The large predicted surface accuracy margin opens up other THz mission applications, beyond BHEX. Ongoing work on a half scale demonstrator model targets TRL 6 by 2027-28.

## 8. Acknowledgements

We gratefully acknowledge A. Lupsasca for supporting the design study and R. Baturin, P. Galison, J. Houston, M. Johnson, and K. Knott for support, encouragement, discussions, and comments in various forms. BHEX is supported by initial funding from Fred Ehrsam.